\documentclass[11pt,letterpaper]{article}

\usepackage{times}
\usepackage{relsize}
\usepackage{microtype}
\usepackage{cite}
\usepackage{xspace}
\usepackage[table,xcdraw]{xcolor}
\usepackage{color,colortbl}
\usepackage{subfigure}
\usepackage{alltt}
\usepackage{url}
\usepackage{comment}
\usepackage{graphicx}
\usepackage{relsize}
\usepackage{paralist}
\usepackage{todonotes}
\usepackage{wrapfig}
\usepackage{titlesec}
\usepackage{epsfig}
\usepackage{geometry}
\usepackage{tikz}
\usepackage[normalem]{ulem}
\usepackage{titlesec}
\usepackage{booktabs}
\usepackage{wrapfig}
\usepackage{multirow}
\usepackage{pifont}
\usepackage{wrapfig}
\usepackage{soul}
\usepackage{enumitem}
\usepackage{tcolorbox}
\usepackage{pdfpages}
\usepackage{fancyhdr}
\usepackage{longtable}
\usepackage{makecell}
\usepackage{lettrine}
\usepackage{listings}
\usepackage{longtable}

\titlespacing{\section}{0pc}{1pc}{1pc}
\titlespacing{\subsection}{0pc}{2pc}{1pc}
\titlespacing{\subsubsection}{0pc}{1pc}{1pc}
\titleformat*{\section}{\Large\bfseries}
\titleformat*{\subsection}{\large\bfseries}
\titleformat*{\subsubsection}{\normalsize\bfseries}

\lstdefinestyle{wcsStyle}{
  tabsize=4,
  showspaces=false,
  showstringspaces=false,
  aboveskip=0em,
  belowskip=0em,
}

\definecolor{Gray}{gray}{0.9}

\newcommand{\pp}[1]{\medskip \noindent \textbf{#1.}\xspace}

\begin{document}

\includepdf[pages=-]{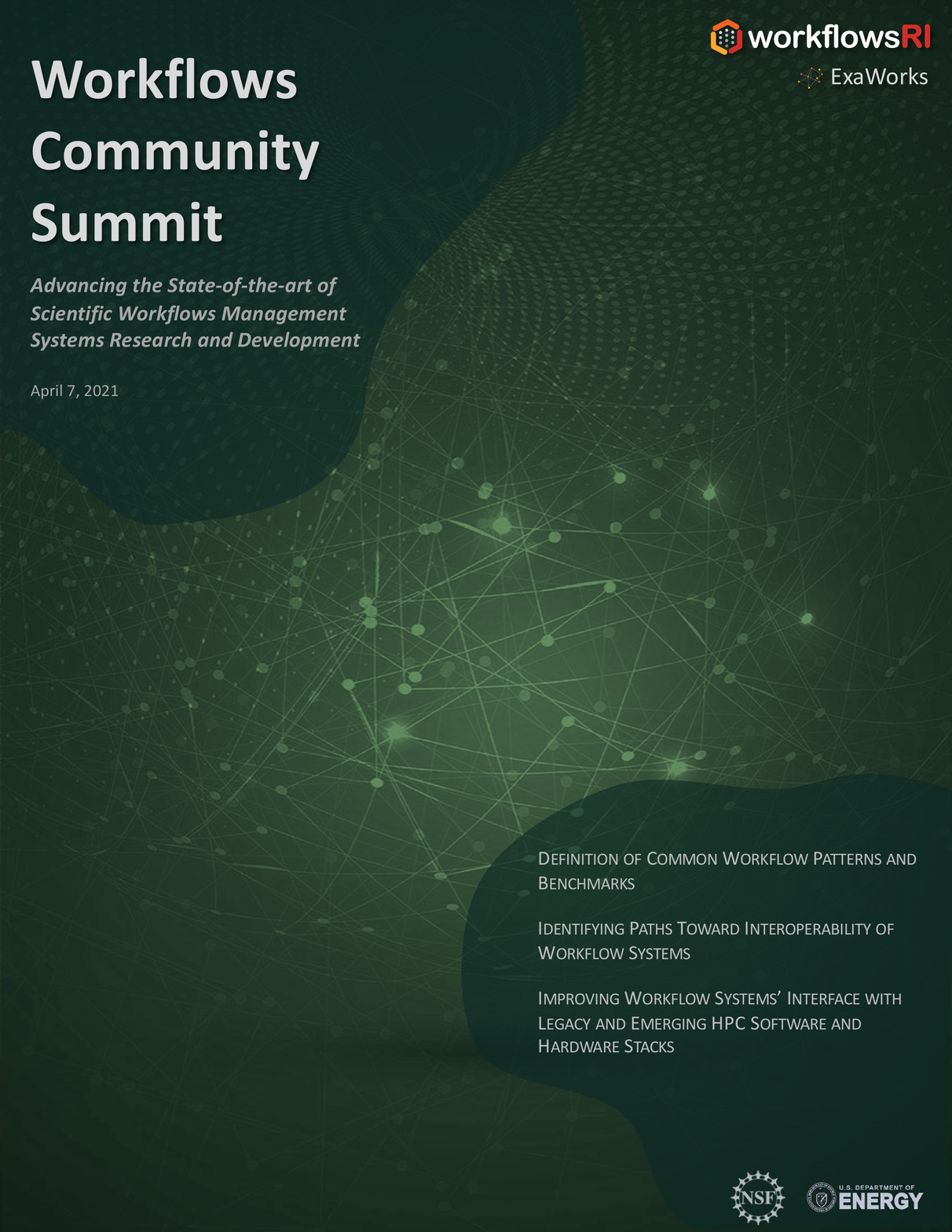}


\pagestyle{fancy}
\fancyhf{}
\rhead{
  \includegraphics[height=12pt]{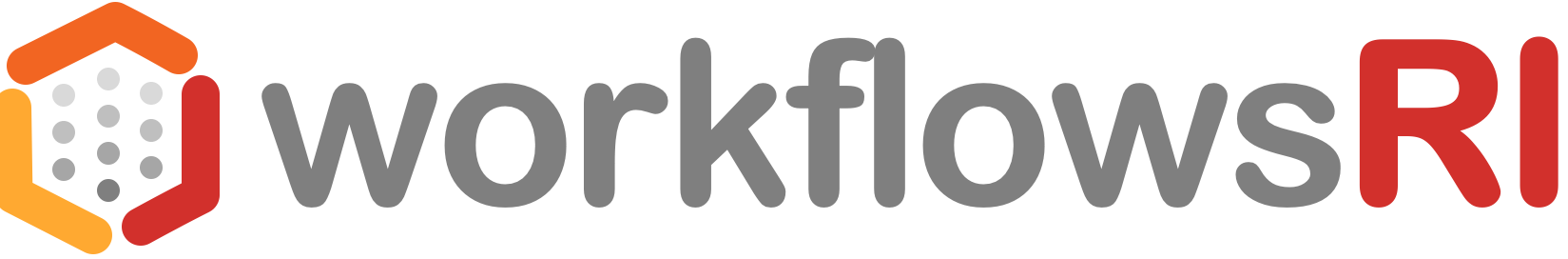} 
  \includegraphics[height=9pt]{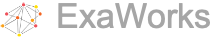}
}
\lhead{Workflows Community Summit - April 2021}
\rfoot{\thepage}


\begin{table}[!ht]
\centering
\smaller
\begin{tabular}{p{16cm}}
    \textbf{Disclaimer}
    \\
    The Workflows Community Summit was supported by the National Science Foundation (NSF) under grants number 2016610, 2016619, and 2016682, and the Department of Energy (DOE). Any opinions, findings, and conclusions or recommendations expressed at the event or in this report are those of the authors and do not necessarily reflect the views of NSF or DOE.
    \\
    \vspace{0.5em}
    \textbf{Preferred citation}
    \\
    R. Ferreira da Silva, H. Casanova, K. Chard, T. Coleman, D. Laney, D. Ahn, S. Jha, D. Howell, S. Soiland-Reys, I. Altintas, D. Thain, R. Filgueira, Y. Babuji, R. M. Badia, B. Balis, S. Caino-Lores, S. Callaghan, F. Coppens, M. R. Crusoe, K. De, F. Di Natale, T. M. A. Do, B. Enders, T. Fahringer, A. Fouilloux, G. Fursin, A. Gaignard, A. Ganose, D. Garijo, S. Gesing, C. Goble, A. Hasan, S. Huber, D. S. Katz, U. Leser, D. Lowe, B. Ludaescher, K. Maheshwari, M. Malawski, R. Mayani, K. Mehta, A. Merzky, T. Munson, J. Ozik, L. Pottier, S. Ristov, M. Roozmeh, R. Souza, F. Suter, B. Tovar, M. Turilli, K. Vahi, A. Vidal-Torreira, W. Whitcup, M. Wilde, A. Williams, M. Wolf, J. Wozniak,
    ``Workflows Community Summit: Advancing the State-of-the-art of Scientific Workflows Management Systems Research and Development", Technical Report, June 2021, DOI: 10.5281/zenodo.4915801.
    \\
    \rowcolor[HTML]{F7F7F7}
    \lstset{basicstyle=\scriptsize,style=wcsStyle}
    \begin{lstlisting}
@misc{wcs2021technical,
  author    = {Ferreira da Silva, Rafael and Casanova, Henri and Chard, Kyle and Coleman,
               Tain\={a} and Laney, Dan and Ahn, Dong and Jha, Shantenu and Howell, Dorran and
               Soiland-Reys, Stian and Altintas, Ilkay and Thain, Douglas and Filgueira,
               Rosa and Babuji, Yadu and Badia, Rosa M. and Balis, Bartosz and Caino-Lores,
               Silvina and Callaghan, Scott and Coppens, Frederik and Crusoe, Michael R. and
               De, Kaushik and Di Natale, Frank and Do, Tu M. A. and Enders, Bjoern and
               Fahringer, Thomas and Fouilloux, Anne and Fursin, Grigori and Gaignard, Alban
               and Ganose, Alex and Garijo, Daniel and Gesing, Sandra and Goble, Carole and
               Hasan, Adil and Huber, Sebastiaan and Katz, Daniel S. and Leser, Ulf and Lowe,
               Douglas and Ludaescher, Bertram and Maheshwari, Ketan and Malawski, Maciej and
               Mayani, Rajiv and Mehta, Kshitij and Merzky, Andre and Munson, Todd and Ozik,
               Jonathan and Pottier, Lo\"{i}c and Ristov, Sashko and Roozmeh, Mehdi and Souza,
               Renan and Suter, Fr\'ed\'eric and Tovar, Benjamin and Turilli, Matteo and Vahi,
               Karan and Vidal-Torreira, Alvaro and Whitcup, Wendy and Wilde, Michael and
               Williams, Alan and Wolf, Matthew and Wozniak, Justin},
  title     = {{Workflows Community Summit: Advancing the State-of-the-art of Scientific
                Workflows Management Systems Research and Development}},
  month     = {June},
  year      = {2021},
  publisher = {Zenodo},
  doi       = {10.5281/zenodo.4915801}
}
    \end{lstlisting}
    \\
    \vspace{0.5em}
    \textbf{License}
    \\
    This report is made available under a Creative Commons Attribution-ShareAlike 4.0 International license ({\scriptsize \url{https://creativecommons.org/licenses/by-sa/4.0/}}).
\end{tabular}
\end{table}
\vspace*{\fill}

\newpage


\cleardoublepage\phantomsection\addcontentsline{toc}{section}{Executive Summary}
\section*{Executive Summary}
\label{sec:summary}

Scientific workflows are a cornerstone of modern scientific computing, and they have underpinned some of the most significant discoveries of the last decade. Many of these workflows have high computational, storage, and/or communication demands, and thus must execute on a wide range of large-scale platforms, from large clouds to upcoming exascale HPC platforms. Workflows will play a crucial role in the data-oriented and post-Moore’s computing landscape as they democratize the application of cutting-edge research techniques, computationally intensive methods, and use of new computing platforms. As workflows continue to be adopted by scientific projects and user communities, they are becoming more complex. Workflows are increasingly composed of tasks that perform computations such as short machine learning inference, multi-node simulations, long-running machine learning model training, amongst others, and thus increasingly rely on heterogeneous architectures that include CPUs but also GPUs and accelerators. The workflow management system (WMS) technology landscape is currently segmented and presents significant barriers to entry due to the hundreds of seemingly comparable, yet incompatible, systems that exist. Another fundamental problem is that there are conflicting theoretical bases and abstractions for a WMS. Systems that use the same underlying abstractions can likely be translated between, which is not the case for systems that use different abstractions.

\subsection*{Summary of the Summit}
\addcontentsline{toc}{subsection}{Summary of the Summit}

The NSF-funded WorkflowsRI project (\url{https://workflowsri.org}) and the DoE-funded ExaWorks project ({\url{https://exaworks.org}}) projects held a very productive Workflows Community Summit in January, 2021 (\url{https://workflowsri.org/summits/community}). The focus was to identify broad challenges for the workflow community and to propose a vision for community activities to address these challenges. While the first summit focused on establishing a high level vision, this second summit explored technical approaches for realizing (part of) that vision. Consequently, we hoped to engage participants who are actively involved in workflow system development. Based on the outcomes of the first summit, we  identified three technical topics for discussion (some align with a single theme of the first summit, while others are cross-cutting): (i)~Definition of common workflow patterns and benchmarks (both for determining workflow system functionality and for providing users with tutorial examples); (ii)~Identifying paths toward interoperability of workflow systems; and (iii)~Improving workflow systems' interface with legacy and emerging HPC software and hardware stacks.

The second edition of the ``Workflows Community Summit" was held online on April 7, 2021. The summit included 75 participants from a group of international researchers and developers (Australia, Austria, Brazil, France, Germany, Italy, Netherlands, Norway, Poland, Spain, Switzerland, UK, and USA) from distinct workflow management systems and users, and representatives from the National Science Foundation~(NSF), and Department of Energy~(DoE). Once again, the summit was co-organized by the PIs of the NSF-funded WorkflowsRI project and the DoE-funded ExaWorks project.

For each of the three technical topics, the lead gave a plenary 5-minute lightning talk followed by focused discussions in breakout sessions. The goal of these sessions was to discussion solutions and technical approaches for achieving the visions for community activities to address broad challenges identified by the workflows community in the previous summit and to develop a ``roadmap" containing deliverables, milestones, action items, etc. (summarized below).

This report documents and organizes the wealth of information provided by the participants before, during, and after the summit. Additional details (including the agenda and presentation videos) can be found at { \url{https://workflowsri.org/summits/technical}}.

\subsection*{Summary of Discussions and Roadmap}
\addcontentsline{toc}{subsection}{Summary of Discussions and Roadmap}

\pp{Topic \#1: Definition of common workflow patterns and benchmarks}
Several of the themes discussed in the previous Workflow Community Summit pointed to strong needs for establishing repositories of common workflow patterns benchmarks, which was part of the discussion for 4 of the 6 themes of the previous summit. In this topic, the goal is to develop workflow patterns with the following requirements: (i)~Each pattern should be easy for users to leverage as starting point for their own specific workflow applications; (ii)~Each pattern should provide links to one or more implementations; and (iii)~Patterns and their implementations should be provided as part of a community-curate repository. The discussion about workflow patterns targeted questions regarding the level of abstraction, the specification of execution platforms and logistics, and existing sources. In the context of workflow benchmarks, the goal is to develop benchmarks with the following requirements: (i)~Specifications should enable workflow developers to develop and determine the support to these specifications; (ii)~Each benchmark should provide links to one or more implementations; and (iii)~These benchmarks should be configurable so that both weak and strong scaling experiments can be conducted. The discussion about workflow benchmarks also targeted the specification of execution platforms and logistics, as well as benchmarks implementation. The following \emph{\textbf{roadmap milestones}} were identified: (i)~Define small sets of workflow pattern and of workflow benchmark deliverables; (ii)~Work with a selected set of workflow system teams to implement the above patterns and benchmarks using their system; (iii)~Investigate options for automatic generation of patterns and/or benchmarks using existing approaches; and (iv)~Identify or create a centralized repository to host and curate the above patterns and benchmarks.

\pp{Topic \#2: Identifying paths toward interoperability of workflow systems}
The previous Workflows Community Summit identified the need to identify paths of interoperability among workflow systems that can happen at multiple technical levels (e.g., task, scheduling, tools, workflows, data, metadata, provenance, and packaging) as well as non-technical such as the semantic level. The need for addressing interoperability issues was part of the discussion for 4 of the 6 themes of the previous summit. The goal of this topic is to discuss several issues related to interoperability of workflow systems and applications. The group has identified the following issues during the discussion: (i)~Lack of interoperability at the software/hardware layers, but more importantly the lack of data interoperability; (ii)~The need for developing horizontal interoperability (i.e., making interoperable components), in addition to vertical solutions; (iii)~The need for separating the abstract workflow from its execution; and (iv)~Lack of sustained funding for supporting interoperability efforts. The following \emph{\textbf{roadmap milestones}} were identified: (i)~Define what interoperability means for different roles including workflow designers, runners, system creators, operators, etc.; (ii)~Establish a requirements document per abstraction layer that will capture the commonalities between components of workflow systems; (iii)~Develop real-world workflow benchmarks featuring different configurations; (iv)~Develop use cases for interoperability based on real-life scenarios; (v)~Develop common APIs that represent a set of workflow library components; and (vi)~Establish a workflow management systems developer community.

\pp{Topic \#3: Improving workflow systems' interface with legacy and emerging HPC software and hardware stacks}
The previous Workflow Community Summit identified the need to focus on improving the interface between workflow systems and existing as well as emerging HPC and cloud stacks. This goal of this topic is to identify challenges faced by workflow systems with respect to discovering and interacting with a diverse set of cyberinfrastructure resources and also the difficulties authenticating remote connections while adhering to facility policies. The group has discussed several important topics including: (i)~Identifying new workflow patterns, e.g. motivated from AI workflows; (ii)~Attaining portability across heterogeneous hardware; (iii)~Developing a registry of execution environment information; and (iv)~Addressing authentication challenges. The following \emph{\textbf{roadmap milestones}} were identified: (i)~Document a machine-readable description of the essential properties of widely used sites; and (ii)~Document remote authentication requirements from the workflow perspective andparticipate in a summit involving workflow system developers, end users, authentication technology providers,and facility operators.

\newpage

\section{Introduction}
\label{sec:introduction}

Scientific workflows are a cornerstone of modern scientific computing, and they have underpinned some of the most significant discoveries of the last decade. Scientific workflows are used to describe complex computational applications that require efficient and robust management of large volumes of data, which are typically stored/processed on heterogeneous, distributed resources. Many of these workflows have high computational, storage, and/or communication demands, and thus must execute on a wide range of large-scale platforms, from large clouds to upcoming exascale HPC platforms~\cite{ferreiradasilva-fgcs-2017}. Workflows will play a crucial role in the data-oriented and post-Moore’s computing landscape as they democratize the application of cutting-edge research techniques, computationally intensive methods, and use of new computing platforms. These discoveries and emerging opportunities are in part a result of decades of workflow management system (WMS) research, development, and community engagement to support the sciences~\cite{atkinson2017scientific}.

As workflows continue to be adopted by scientific projects and user communities, they are becoming more complex. Workflows are increasingly composed of tasks that perform computations such as short machine learning inference, multi-node simulations, long-running machine learning model training, amongst others, and thus increasingly rely on heterogeneous architectures that include CPUs but also GPUs and accelerators.
As a result of widespread workflow adoption, the workflow research and development community has grown: there are now hundreds of independent WMSs~\cite{workflow-systems}, thousands of researchers and developers, and a rapidly growing corpus of workflows research publications. The WMS technology landscape is thus segmented and presents significant barriers to entry due to the hundreds of seemingly comparable, yet incompatible, systems that exist. Another fundamental problem is that there are conflicting theoretical bases and abstractions for a WMS. Systems that use the same underlying abstractions can likely be translated between, which is not the case for systems that use different abstractions.

The NSF-funded WorkflowsRI project~\cite{workflowsri} and the DoE-funded ExaWorks project~\cite{exaworks} projects held a very productive Workflows Community Summit in January (\url{https://workflowsri.org/summits/community}). The focus was to identify broad challenges for the workflows community and to propose a vision for community activities to address these challenges. While the first summit focused on establishing a high level vision, this second summit explored technical approaches for realizing (part of) that vision. Consequently, engaged participants who are actively involved in workflow system development. Based on the outcomes of the first summit, we identified three technical topics for discussion (some align with a single theme of the first summit, while others are cross-cutting): 

\begin{compactenum}
    \item Definition of common workflow patterns and benchmarks (both for determining workflow system functionality and for providing users with tutorial examples);
    \item Identifying paths toward interoperability of workflow systems; and
    \item Improving workflow systems' interface with legacy and emerging HPC software and hardware stacks.
\end{compactenum}

\subsection{Summit Organization}

\begin{figure}[!t]
    \centering
    \includegraphics[width=\linewidth]{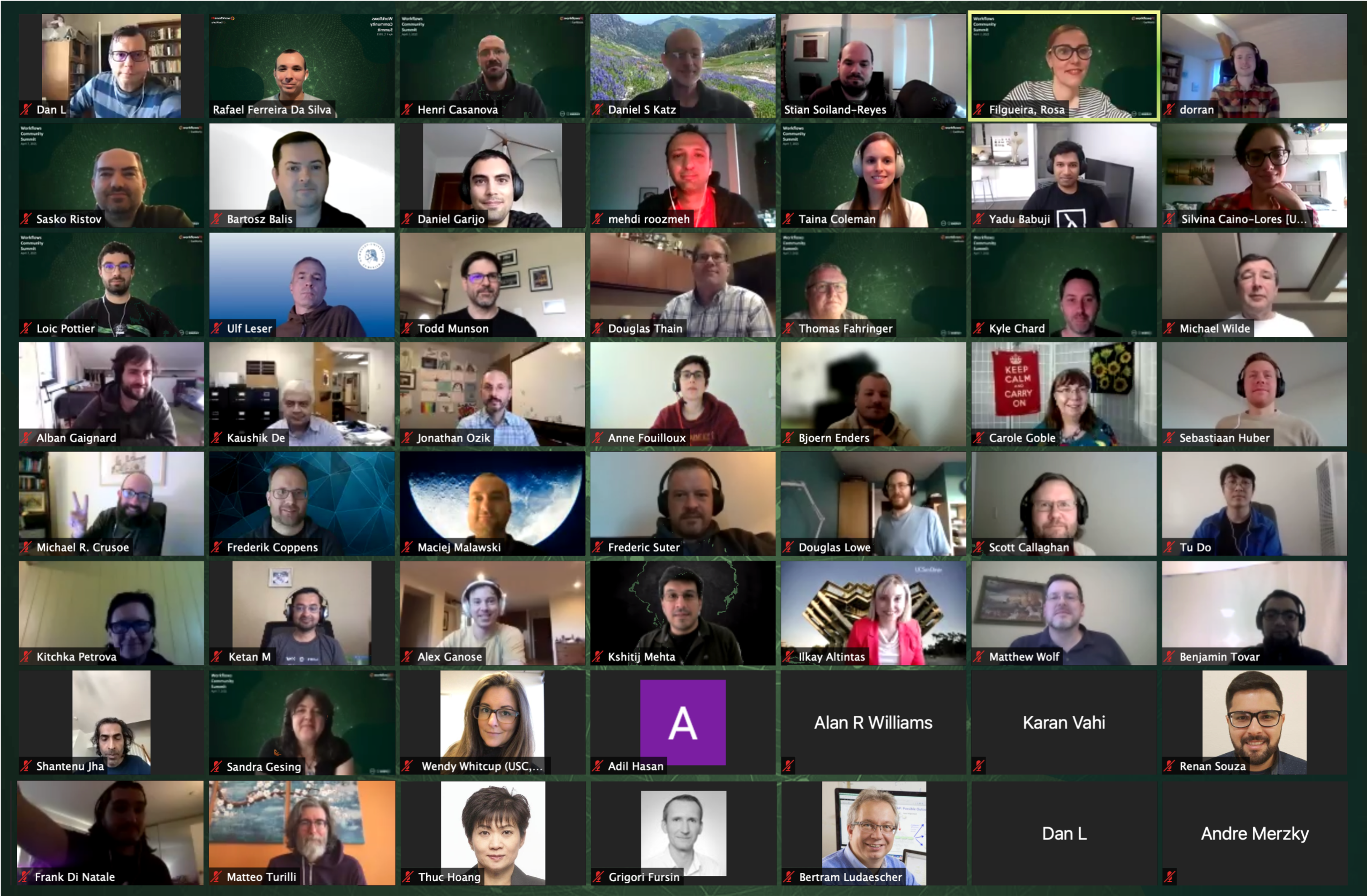}
    \caption{Screenshot of the second edition of the Workflows Community Summit participants. (The event was held virtually via Zoom on April 7, 2021.)}
    \label{fig:participants}
\end{figure}

This document reports on discussions and findings from the second edition of an international ``Workflows Community Summit" that took place on April 7, 2021~\cite{wcs}. 
The summit included 75 participants from a group of international researchers and developers (Australia, Austria, Brazil, France, Germany, Italy, Netherlands, Norway, Poland, Spain, Switzerland, UK, and USA)  from distinct workflow management systems and users, and representatives from the US National Science Foundation (NSF), and the US Department of Energy (DoE) (Figure~\ref{fig:participants}).

\begin{table}[!t]
    \centering
    \small
    \setlength{\tabcolsep}{12pt}
    \begin{tabular}{lll}
        \toprule
        \textbf{Project} & \textbf{Name} & \textbf{Affiliation} \\
        \midrule
        WorkflowsRI & Rafael Ferreira da Silva & University of Southern California \\
        & Kyle Chard & University of Chicago \\
        & Henri Casanova & University of Hawai'i at M\~anoa \\
        & Tain\~a Coleman & University of Southern California \\
        \rowcolor[HTML]{F2F2F2}
        ExaWorks & Dan Laney & Lawrence Livermore National Laboratory \\
        \rowcolor[HTML]{F2F2F2}
        & Dong H. Ahn & Lawrence Livermore National Laboratory \\
        \rowcolor[HTML]{F2F2F2}
        & Kyle Chard & Argonne National Laboratory\\
        \rowcolor[HTML]{F2F2F2}
        & Shantenu Jha & Brookhaven National Laboratory \\
        \bottomrule
    \end{tabular}
    \caption{Workflows Community Summit organizers.}
    \label{tab:organization}
\end{table}

The summit was co-organized by the PIs of two distinct projects: the NSF-funded WorkflowsRI project~\cite{workflowsri} and the DoE-funded ExaWorks project~\cite{exaworks} (Table~\ref{tab:organization}). These two projects aim to develop solutions to address some of the challenges faced by the workflow community, albeit in different ways. 

\pp{WorkflowsRI}
Focuses on defining the blueprints for a community research and development infrastructure. Specifically, the goal is to bring together workflow users, developers, and researchers to identify common ways to develop, use, and evaluate workflow systems, resulting in a useful, maintainable, and community-vetted set of principles, methods, and metrics for WMS research and development. The resulting research infrastructure will provide an academic nexus for addressing pressing workflow challenges, including those resulting from the increasing heterogeneity and size of computing systems, changing workflow patterns to encompass rapid feedback and machine learning models, and the increasing diversity of the user community. 

\pp{ExaWorks} 
Aims to develop a multi-level workflows SDK that will enable teams to produce scalable and portable workflows for a wide range of exascale applications. ExaWorks does not aim to replace the many workflow solutions already deployed and used by scientists, but rather to provide a robust SDK and work with the community to identify well-defined and scalable components and interfaces that can be leveraged by new and existing workflows. A key goal is that SDK components should be designed to be usable by many other WMS, thus facilitating software convergence in the workflows community. Although clearly distinct, these two projects have similar motivations, and their planned research and development activities could benefit from collaboration and even co-design.

\subsection{Summit Structure and Activities}

Based on the outcomes of the first summit, we identified six cross-cutting technical topics for the second edition of the summit, each of which was the object of a focused discussion led by a volunteer community member (Table~\ref{tab:lead}). For each of the technical topics shown in Table~\ref{tab:lead}, the lead gave a plenary 5-minute lightning talk followed by focused discussions in breakout sessions. The goal of these sessions was to discussion solutions and technical approaches for achieving the visions for community activities to address broad challenges identified by the workflows community in the previous summit and to develop a ``roadmap" containing deliverables, milestones, action items, etc. The lead then reported on the outcome of the discussion in plenary sessions, after which final remarks were given by the organizers and the summit was adjourned. In the following sections, we summarize the outcome from the breakout session for each theme.

All presentations and videos can be found in the summit website (\url{https://workflowsri.org/summits/technical/}), and videos can be watched from the WorkflowsRI's YouTube channel (\url{https://www.youtube.com/watch?v=lbGCz2EgfZU&list=PLAtmuqHExRvOfsFTDdtGDrhovi1oQWLgP}).

\begin{table}[!t]
    \centering
    \small
    \begin{tabular}{p{8.5cm}|l}
        \toprule
        \textbf{Topic} & \textbf{Discussion Co-Leaders} \\
        \midrule
        Topic \#1: Definition of common workflow & Dan Laney,  Lawrence LLNL, USA \\
        patterns and benchmarks & Dorran Howell, Tweag I/O, France \\
        \rowcolor{Gray}
        Topic \#2:  Identifying paths toward interoperability of & Stian Soiland-Reyes, Univ. of Manchester, UK \\
        \rowcolor{Gray}
        workflow systems & Ilkay Altintas, UC San Diego, USA \\
        Topic \#3: Improving workflow systems' interface with & Douglas Thain, Univ. of Notre Dame, USA \\
        legacy and emerging HPC software and hardware stacks & Rosa Filgueira, Heriot-Watt University, UK \\
        \bottomrule
    \end{tabular}
    \caption{Workflows Community Summit topics and discussion co-leaders.}
    \label{tab:lead}
\end{table}

\newpage
\section{Summit Topics}

\subsection{Topic \#1: Definition of common workflow patterns and benchmarks}
\label{sec:topic1}


Several of the themes discussed in the previous Workflow Community Summit~\cite{ferreiradasilva2021wcs} pointed to strong needs for establishing repositories of common workflow patterns benchmarks.  Specific findings from the previous Summit are summarized in Section~\ref{sec:topic1:findings}. Given these findings, the discussion during the breakout session of Topic~\#1 in this Workflow Community Summit focused on defining two broad sets of deliverables that speak to the above needs: (i)~Workflow patterns; and (ii)~Workflow benchmarks.  Both kinds of deliverables are described in detail in Sections~\ref{sec:topic1:deliverable1} and~\ref{sec:topic1:deliverable2}. Finally, the discussion during the breakout session defined a roadmap with possible list of milestones and action items, which are given in Section~\ref{sec:topic1:milestones}.

\subsubsection{Findings from previous summit}
\label{sec:topic1:findings}

The need for common workflow patterns and benchmark was part of
the discussion for 4 of the 6 themes of the previous summit:

\pp{Workflow patterns}
A large part of the \emph{training and education for workflow
users} (Theme \#2) discussion focused on workflow patterns with two main
goals: (i)~reduce the barrier of entry for prospective users; and
(ii)~enable transfer of expertise when adopting new workflow systems.
Workflow patterns were also discussed in the context of \emph{exascale
challenged and beyond} (Theme \#4). Given the technical intricacies of
workflow executions on HPC platforms, it was determined that workflow
patterns that users could inspect and or modify for their own purposes
would be of great value and serve as effective documentation.  Finally,
workflow patterns were discussed in the context of \emph{API,
interoperabilty, reuse, and standards} (Theme \#5), the idea being that
workflow patterns would be a good way to expose or discover commonalities
between workflow systems and their APIs.

\pp{Workflow benchmarks}
A large part of the \emph{exascale challenges and beyond}
(Theme \#4) discussion focused on workflow benchmarks, the goal being to
not only benchmark workflow systems on HPC platforms, but also to include
these workflow benchmarks in the machine procurement and acceptance test
processes.  Benchmarks were also discussed in the context of \emph{AI
Workflows} (Theme \#3), with the goal of defining benchmarks representative
of common workflow patterns for AI applications, which are of
critical importance and require the large amounts of compute resources
provided by HPC platforms. In particular, AI workflows benefit immensely
from exploiting hybrid CPU-GPU architectures that are now common-place in
high-end HPC platforms.

\subsubsection{Workflow patterns}
\label{sec:topic1:deliverable1}

\pp{Requirements} 
The goal is to develop workflow patterns with the following requirements:

\begin{compactenum}

    \item Each pattern should be easy for users to leverage as starting point for their own specific workflow applications, or as a way to learn how to use particular workflow systems.

    \item Each pattern should provide links to one or more implementations, where each implementation is for a particular workflow system and can be downloaded and modified easily by users. These implementations could be provided, maintained, and evolved by workflow system developers who want to advertise their systems' capabilities to the workflow user community. They could also be provided by groups of users who have successfully implemented the pattern using a workflow system. Implementations should be documented and packaged using common practices (e.g., common documentation formats and structure, containers, testing framework).  Although this may be difficult to achieve, ideally implementations would also provide testbed compute resources for execution (e.g., via web portals).

    \item Patterns and their implementations should be provided as part of a community-curated GitHub or equivalent repository.

\end{compactenum}

\pp{Question: Level of abstraction} 
A broad question that arose during the discussion and that will require further consultation and community-involvement is that of the \emph{level of abstraction} of the workflow patterns, that is the level of
connection to real application use-cases -- At one extreme, workflow patterns could be completely abstract with no connection to any real-world application. For instance static workflow patterns could include ``a fork-join pattern with data depencies via files"; and dynamic workflow patterns could include ``a loop that generates 10 independent tasks at each iteration and then creates 10 new independent tasks based on the output of the previous 10 tasks".  At the other extreme, workflow patterns could be completely use-case driven and correspond to actual scientific applications, with realistic task computations and datasets.  Purely abstract workflow patterns are likely useful for tutorial (``hello world") purposes, and should probably be included regardless. Use-case patterns would also be useful when curated, so that they are organized by scientific domains and many patterns are included for each domain. The goal is to identify useful patterns that span the spectrum of possible levels of
abstraction.  The discussion also touched on best practices for new workflow users who are not necessarily used to ``workflow thinking". It was mentioned that a possible way to have users move from being script-authors to workflow-authors is to use simple models and annotation languages, such as YesWorkflow~\cite{yes-workflow}, which would be used to develop a minimum set of intro-level patterns for users with no previous workflow exposure.

\pp{Question: Specification of execution platforms and logistics} 
The question here is the level of detail with which a pattern specifies the platform on which it is to be executed  and the logistics of the execution on that platform. The platform description could be left completely abstract, e.g., ``a set of compute locations and storage locations interconnected by some network"; or it could be fully specified, e.g., ``a set of 10 AWS instances with these specifications and this kind of cloud storage, with these specific software systems installed with these configuration files"  Perhaps it is only necessarily to define broad classes of execution settings, e.g., HPC vs. cloud environments.  Regarding the logistics of the executing platform, the pattern could specify very little, or could specify full details (e.g., file and directory management, chaining and statusing, data flow mechanisms between tasks, etc.). Under-specifying execution platforms and logistics may render the pattern not useful, but over-specifying them could render the pattern too niche.

\pp{Question: Existing sources} 
Another question was that of which existing sources of workflow patterns/motifs could be leveraged for deliverable relevant workflow patterns. The following specific such sources were mentioned by participants:

\begin{compactitem}
    \item Van der Aalst work~\cite{russell2016workflow} on workflow patterns based on common control flow, branching and state patterns \url{http://workflowpatterns.com}, \url{http://mlwiki.org/index.php/Workflow_Patterns}
    \item Garijo and Alper work on common workflow motifs based on their functionalities and common operations \cite{garijo2014common}. Expanded work can be seen in their respective theses \cite{garijo2015mining, alper2016towards}.
    \item Workflow repositories useful for mining purposes, such as myExperiment~\cite{goble2010myexperiment} (\url{https://www.myexperiment.org/workflows}) and Workflow Hub (\url{https://workflowhub.eu/)}.
    \item Synthetic workflows and generators (\url{https://wfcommons.org})~\cite{coleman2021wfcommons}.
    \item Early/high-level work comparing scientific and business workflows~\cite{ludascher2009scientific} .
    \item Starlinger work on workflows similarity~\cite{starlinger2016similarity}.
\end{compactitem}

\subsubsection{Workflow benchmarks}
\label{sec:topic1:deliverable2}

\pp{Requirements}
The goal is to develop workflow benchmarks with the following requirements:

\begin{compactenum}

    \item Benchmark specifications should make it easy for workflow developers to develop them or to determine that their system cannot implement these specifications.

    \item  Each benchmark should provide links to one or more implementations, where each implementation is for a particular workflow system. These implementations would be provided, maintained, and evolved by workflow system developers. They should be able to be packaged so that they are executed out of the box on the classes of platforms they support.

    \item These benchmarks should be configurable so that they scale for different input sizes, so that both weak and strong scaling experiments can be conducted.

\end{compactenum}

\pp{Question: Specification of execution platforms and logistics}
This is the same question as that in the previous question, as it applied both to workflow patterns and workflow benchmarks. It was discussed for both patterns and benchmarks simultaneously during the Summit.

\pp{Question: Benchmark implementation} 
A large part of the discussion focused on ways in which these benchmarks should be developed. Options considered included: real-world applications, mini-apps (also called proxy-apps), benchmarks generated based on application skeletons~\cite{katz2016application}, and automatically generated benchmarks based on synthetic workflow generators~\cite{da2020workflowhub, wfcommons}.  While automatic generation is obviously attractive, it was noted that it can be fraught with complexity. As a case in point, the use of application skeletons for this purpose can even require the development of a specific language (as done in~\cite{katz2016application}).

\subsubsection{Roadmap}
\label{sec:topic1:milestones}


The following milestones were identified by the end of the discussion:

\begin{compactitem}

    \item Define small sets (say between 5 and 10) of workflow pattern and of workflow benchmark deliverables, based on the requirements and tackling the questions in the previous two sections. These should be defined by eliciting feedback from users and workflow system developers, as well as based on existing sources that provide or define real-world or synthetic workflow patterns. The hope is that
        good coverage of needs can be achieved provided these small sets of patterns and benchmarks are picked judiciously.


    \item Work with a selected set of workflow system teams to implement the above patterns and benchmarks using their system.

    \item Investigate options for automatic generation of patterns and/or benchmarks using existing approaches (application skeletons, synthetic workflow generators, etc.)

    \item Identify or create a centralized repository to host and curate the above patterns and benchmarks.

\end{compactitem}

\subsection{Topic \#2: Identifying paths toward interoperability of workflow systems}
\label{sec:topic2}

There is a number of workflow management systems~\cite{workflow-systems} that differ at varying degrees such as expressivity, execution models, and ecosystems. These differences are mainly due to individual implementations of language, control mechanisms (e.g., fault tolerance, loops), data management mechanisms, execution backends, reproducibility aspects for sharing workflows, and provenance and FAIR metadata capturing. The previous Workflows Community Summit~\cite{ferreiradasilva2021wcs} identified the need to identify paths of interoperability among these systems that can happen at multiple technical levels (e.g., task, scheduling, tools, workflows, data, metadata, provenance, and packaging) as well as non-technical such as the semantic level -- need an agreement on the terminology. Additional non-technical challenges also include organizational and legal issues (e.g., licenses compatibility, data sharing policies). This breakout group focused on the challenges faced by the lack of interoperability for designing workflow systems and/or executing workflows. The goal was to identify layers of interoperability that could be practically achievable and related efforts that could be fostered. Appendix~A provides an example list of efforts for different interoperability layers including technical, semantic, organizational, and legal workflow interoperability.

\subsubsection{Findings from previous summit}

The need for addressing interoperability issues was part of the discussion for 4 of the 6 themes of the previous summit:

\pp{Standards (Theme \#5)}
Each workflow system serves a different user community or underlying compute engine, thus comparing the differences and commonalities between these systems is of paramount importance. A first step would be to identify and characterize domain-specific efforts, and develop case studies of, for example, business process workflows and serverless workflow systems for identifying workflow patterns. To achieve these goals, sustained funding for developing workflow standards is fundamental. 

\pp{FAIR Computational Workflows (Theme \#1)}
The FAIR principles~\cite{wilkinson2016fair} have laid a foundation for sharing and publishing digital assets and in particular data. The challenge is how to design workflows in a way they are interoperable across systems, how to automate FAIRness in workflows, and even enable interoperability among workflow systems themselves. Such workflows should be properly documented with the necessary provenance and metadata information, and shared via FAIR workflow repositories. More details on the adapting the FAIR principles for computational workflows can be found in~\cite{goble_2020}.

\pp{Heterogeneous computing (Theme \#3)}
Scientific workflows empowered with machine learning (ML) techniques are mainstream in scientific computing nowadays. A key interoperability challenge is how to support for heterogeneity of compute hardware (e.g., CPU/GPU).

\pp{Optimizing for exascale (Theme \#4)}
In an attempt to optimize the execution of applications on the forthcoming exascale systems, developers tend to focus on low level (and often hard coded) optimizations. These optimizations often cannot be easily/fully propagated to other systems, thus impairing the path to interoperability.

\subsubsection{Discussion}

The group discussed several issues related to interoperability of workflow systems and applications.

\pp{Lack of interoperability}
The need for interoperability of workflow applications and systems is commonly modeled as a problem of porting applications across systems, which may require days up to weeks of development effort, e.g. porting from Snakemake on HPC to Hadoop on a commodity cluster using YARN~\cite{schiefer2020portability}. In addition to the software/hardware component, the group has also identified the lack of data interoperability: how to access data as a high-level abstraction? The current state of the art in workflow representation provides sophisticated mechanisms for representing connectivity among tasks within a workflow, however data is still considered a second class artifact -- i.e., there is not a common API or an abstraction to access/manage data across different storage resources. Last, the lack of interoperability has also been characterized by the inability to ``think ahead". This topic clearly affects the interplay of workflow systems and stable Research Data Management.

\pp{Vertical \emph{vs.} horizontal interoperability}
Most of the previous approaches for tackling the interoperability problem attempted to develop complete vertical solutions. For instance, the community has proposed a number of frameworks that are primarily focused on establishing these solutions as the ``one-size-fit-all" product -- with opportunities to subsume other workflow systems. However, there is no attempt to develop an approach from a perspective of making interoperable components. For instance, could we simply replace the scheduler of WMS with the one from another WMS (of course not), or the resource manager of a runtime system with the resource manager of another runtime system (of course not)? Additionally, interoperable components require standardized APIs, which is still an open challenge. There are some solutions (e.g.,~\cite{turilli2019middleware, billings2017toward}) in which performance and a building block approach have been combined -- interoperability is built as a prerequisite to engineering an end-to-end workflow middleware stack. Related efforts include a taxonomy of workflow characteristics~\cite{liew2016scientific} and the design considerations for workflow management systems~\cite{ahmed2021design}. One might also be able to learn from standardization approaches in the middleware world in the 90ties (e.g., CORBA services~\cite{bastide2000formal}).

\pp{Entwining model of execution and data structures}
There is a tendency to tie the workflow with the model of execution and data structures. As a result, it is hard to separate the abstract workflow from its execution. Additionally, there is a need for understanding which component in the architecture of a WMS is responsible for which functionality (e.g., scheduling, data access patterns, graph structure, etc.). Therefore, separation of concerns is crucial for interoperability at many levels, including separation of orchestration of the workflow graph from its execution.

\pp{Funding avenues}
Sustained funding is key for supporting interoperability efforts. Although ad-hoc efforts may have already resulted in initial discussions around the topic, the lack of a more systematic support has significantly impaired further development of the proposed solutions. A potential approach would be to adopt a ``proposal-model" (as in the EOSC-Life project~\cite{eosc-life}) in which researchers can submit proposals for subfunding, e.g. ``Develop an API". In addition to financial support, there is also a need for a generalizable pipeline for ``research to practice"~\cite{abramson2019translational} that would drive workflow research outcomes into community-developed practical tools.

\subsubsection{Roadmap}

The following deliverables/milestones were identified by the end of the discussion:

\begin{compactitem}

    \item Define what interoperability means for different roles: workflow designers, workflow runners, workflow system creators, operators, etc.

    \item Establish a ``requirements" document per \emph{abstraction layer} that will capture the commonalities between components of workflow systems. The goal is to clearly perform a separation of concerns to identify interopeability gaps per layer. A first deliverable would include the demonstration of interoperability for a single layer, i.e., the ability to replace a certain layer in a given WMS with an implementation of this layer from another system.
    
    \item Develop real-world workflow \emph{benchmarks} featuring fixed input types but with different sizes, and/or fixed parameters for checking systems with multiple workflows (of increasing complexity, e.g. linear, DAG, recursive). Such benchmarks would be key to evaluate the efficiency of workflow systems and computing platforms systematically.
    
    \item Develop \emph{use cases} for interoperability based on real-life scenarios, for example, a scientist that is developing their workflows experiments using notebooks and need to port that same experiment across platforms. Another example is the re-execution of a workflow developed in one data center using a certain file system and resource manager to another data center with a different file system and / or different resource manager.
    
    \item Develop \emph{common APIs} that represent a set of workflow library components, so as interoperability could be achieved at the component level (in contrast to interoperability at the workflow specification level as targeted, for instance, by CWL~\cite{cwl}). Efforts such as the Common Workflow Language (CWL)~\cite{cwl}, IWIR Metaworkflows~\cite{arshad2015definition} and Collective Knowledge (CK)~\cite{Fursin_2021}. These common APIs also need to provide APIs for defining inputs, storing intermediate results, and output data.
    
    \item Establish a workflow management systems \emph{developer community}. An immediate actionable item would be to develop a centralized repository of workflow-related research papers, and a workflow system registry aimed at DevOps and/or users. Such repository could leverage text mining methods to categorize and cluster workflow efforts (from research papers).
    
\end{compactitem}

\subsection{Topic \#3: Improving workflow systems' interface with legacy and emerging HPC software and hardware stacks }
\label{sec:topic3}


The previous Workflow Community Summit~\cite{ferreiradasilva2021wcs} identified the need to focus on improving the interface between workflow systems and existing as well as emerging HPC and cloud stacks. This challenge is particularly important as workflows are designed to be used for long periods of time and may be moved between computing providers. Increasingly specialization of hardware and software systems (e.g., with accelerators, virtualization, and cloud infrastructure). This breakout group focused on the challenges faced by workflow systems with respect to discovering and interacting with a diverse set of cyberinfrastructure resources and also the difficulties authenticating remote connections while adhering to facility policies.

\subsubsection{Findings from previous summit}

The previous summit identified several important challenges facing modern workflow systems.  We briefly highlight these challenges here and refer to them throughout the section. 

\pp{Complex workflows (Theme \#3)}
Modern scientific workflows, particularly in machine learning, have components that may perform better on specialized hardware (e.g., CPU, GPU, TPU). The complexity is derived from the challenges  to determine and assess resource requirements throughout the workflow and matching those requirements to the increasingly diverse hardware availability. 

\pp{Diverse cyberinfrastructure (Theme \#4)}
Cyberinfrastructure resources are becoming increasingly diverse, not only in terms of CPU architectures but also in the range of accelerators and layouts of nodes. This diversity promises greater performance, however, at present descriptions are embedded in online documentation that is inaccessible to workflows (and users in many cases). Further, allocation policies and schedulers are not designed with workflows in mind. Typically, they provide relatively coarse grained (in terms of both space and time) allocation units and have unpredictable queue delays. Many workflows require more precise control over resources and job behavior.

\pp{Interoperability (Theme \#5)} 
Increasing software and hardware specialization is further contributing to the challenges of reusing existing  workflow systems and the likelihood of users developing their own workflow tools. One approach for reducing this challenge is to consider interoperability between workflow components, enabling existing workflow systems to leverage components that are developed to work with specialized usage patterns or hardware configurations.

\subsubsection{Discussion}

The group focused on several important topics that considered new workflow patterns, changing hardware and portability, developing a registry of information, and addressing authentication challenges.

\pp{New workflow patterns}
While most workflow patterns have been well-understood for some time, there may be some new requirements emerging. Some of which are motivated from AI workflows, but many are not unique to AI workflows.  Examples of these requirements include: non-determinism with respect to workflows and components of those workflows, use of end-to-end and online optimization techniques to guide decisions (e.g., scheduling, resource provisioning), and the increasing need for heterogeneous hardware for different workflow components. 

\pp{Hardware and portability}
Portability is often stated as a reason for adopting workflow solutions and workflow users are increasingly eager to move to new hardware to take advantage of performance or capabilities. However, in practice portability is challenging and the increasing heterogeneity of hardware will likely make portability both increasingly necessary and also more difficult. The group identified several key challenges that make it difficult to move between systems, including custom and site-specific schedulers, custom hardware,  hand-tuned configurations and packages, hard-coded paths, and various other site-specific assumptions. 

\pp{Registries}
In order for workflow systems to ``intelligently" place and adapt applications for a specific execution environment, they must understand properties of both the application and the underlying systems.  For example, it is important to understand the hardware, authentication system, file system, module system, scheduler type. Further, facilities have specific usage policies that should be followed by users and workflow systems; however, these policies are often ignored due to a lack of knowledge or enforcement. Much of the information needed by workflow systems remains constant at a site and thus could be collected and made available to workflow systems. Currently, individual users and sometimes workflow systems do this independently, establishing their own formats and maintaining some information about different sites.  A potential approach might be to develop a cascading-like model in which different levels of information are layered upon one another for users. For example, facilities and federation (e.g., XSEDE) could describe their resources and policies for using resources; workflow management systems may define appropriate ways of using those resources; and users may have further information that is specific to their use (e.g., accounts, username). 

\pp{Authentication}
Authentication has long been a challenge for workflow systems and  remote computing in general. SSH keys enabled remote access, allowing users to delegate long running agents. GSISSH improved upon this model by enabling delegating authorization via proxy certificates. However, two-factor authentication makes automation difficult as it requires explicit user actions during the authentication process. The web has standardized on OAuth for similar needs, and we seem similar adoption in research computing with several facilities now offering OAuth-based access and Globus using OAuth for authentication with the  Globus Connect software deployed on storage systems. The OAuthSSH (\url{https://github.com/XSEDE/oauth-ssh}) system provides support for OAuth-based authentication via SSH following a model similar to GSISSH. However, these efforts do not remove the need for human approval in automated deployments, rather they provide a secure model for granting approval for a short period of time.  The need for programmatic access to computing facilities is much more broad than workflows and includes science gateways, web services, and other science applications. Thus, there needs to be a broad community approach between HPC administrators, system developers, and users to understand the most suitable models for authentication in automated processes.

\subsubsection{Roadmap}

The breakout group identified the following research activities:

\pp{Registry of site capabilities}
Workflow systems require a standard method for querying a site for information about how to use that site, for example, information about the batch system, file system configuration, data transfer methods, and machine capabilities. Prior efforts, particularly in Grid computing, explored various catalog-like models for capturing and sharing such information (e.g., Globus MDS, GLUE, and OSG catalogs). A crucial first step in this effort is to first understand what information is needed by workflow systems, what information could be made available automatically and what would need to be manually curated, and to understand prior approaches and determine if they can be adapted and leveraged for this purpose. A summit between workflow system developers and facility representatives may be the best way to determine an appropriate model and responsibility for accuracy of the information. There is an ongoing effort by the Science Gateways Community Institute and XSEDE to represent computational resources (\url{https://github.com/SGCI/sgci-resource-inventory}) which may be a good starting point for this effort.

\uline{Expected outputs:} Document a machine-readable description of the essential properties of widely used sites. For example, a first step is to define a JSON schema and share it in GitHub.

\pp{Workflow component authentication}
Remote job execution is an important capability provided by workflow management systems. Remote execution is necessary in cases where workflows span facilities, to avoid deploying workflow management software on login nodes, and to improve user experience (e.g., by enabling users to drive workflows from their laptops). However, authentication has always been  challenging. Many workflow systems rely on fragile SSH connections and in the past the use of GSISSH for delegated authentication. Recently,  sites have moved towards two factor authentication and even OAuth-based solutions. There are also ongoing efforts to provide programmatic identity and access management in scientific domains, two prominent examples are Globus Auth and SciTokens.  While the topic of remote authentication is much more broad than the workflows community, there are important considerations that should be included in this discussion related to programmatic access, community credentials, and long-term access. 

\uline{Expected outputs:} Document remote authentication requirements from the workflow perspective and participate in a summit involving workflow system developers, end users, authentication technology providers, and facility operators.

\newpage
\cleardoublepage\phantomsection\addcontentsline{toc}{section}{References}
\bibliographystyle{IEEEtran}
\bibliography{references}

\newpage
\cleardoublepage\phantomsection\addcontentsline{toc}{section}{Appendix A: Interoperability Layers}
\section*{Appendix A: Interoperability Layers}
\label{appx:interoperability}

\begingroup
\smaller
\begin{longtable}[!ht]{lll}
    \toprule
    \multicolumn{3}{l}{\bf Technical Workflow Interoperability} \\
    \midrule
    Execution & DRMAA & \url{http://www.drmaa.org} \\
    & GA4GH APIs (TRS, WES) & \url{https://www.ga4gh.org/genomic-data-toolkit} \\
    & OGF (JSDL, OGSA) & \url{https://www.ogf.org/ogf/doku.php/standards/standards} \\
    
    \rowcolor[HTML]{F2F2F2}
    Data access & S3 & \url{https://aws.amazon.com/s3} \\
    \rowcolor[HTML]{F2F2F2}
    & DRS & \url{https://ga4gh.github.io/data-repository-service-schemas} \\
    \rowcolor[HTML]{F2F2F2}
    & GridFTP & \url{https://www.ogf.org/documents/GFD.20.pdf} \\
    
    Syntactic & CWL & \url{https://www.commonwl.org} \\ 
    & OpenWDL & \url{https://openwdl.org} \\
    
    \rowcolor[HTML]{F2F2F2}
    Packaging & RO-Crate & \url{https://www.researchobject.org/ro-crate/1.1/workflows.html} \\
    \rowcolor[HTML]{F2F2F2}
    & BioConda & \url{https://bioconda.github.io} \\
    \rowcolor[HTML]{F2F2F2}
    & BioContainers & \url{https://biocontainers.pro/} \\
    \rowcolor[HTML]{F2F2F2}
    & Debian-Med & \url{https://www.debian.org/devel/debian-med} \\
    \rowcolor[HTML]{F2F2F2}
    & Collective Knowledge & \url{https://github.com/ctuning/ck} \\
    
    Data & HDF5 & \url{https://support.hdfgroup.org/HDF5} \\
    & VOTable (IVOA) & \url{https://www.ivoa.net/documents/VOTable} \\
    & CSV On the Web & \url{https://www.w3.org/TR/tabular-data-primer} \\
    & SBML (COMBINE) & \url{https://synonym.caltech.edu} \\
    & HL7 FHIR & \url{https://www.hl7.org/fhir} \\
    & DFDL & \url{https://www.ogf.org/documents/GFD.240.pdf} \\
    & ADIOS & \url{http://adios.ornl.gov} \\
    & Arrow & \url{https://arrow.apache.org} \\
    
    \rowcolor[HTML]{F2F2F2}
    Metadata & DCAT2 & \url{https://www.w3.org/TR/vocab-dcat-2} \\
    \rowcolor[HTML]{F2F2F2}
    & Codemeta & \url{http://codemeta.github.io} \\
    \rowcolor[HTML]{F2F2F2}
    & Datacite & \url{https://schema.datacite.org} \\
    \rowcolor[HTML]{F2F2F2}
    & schema.org & \url{https://schema.org} \\
    
    Repository & WfCommons & \url{https://wfcommons.org} \\
    & WorkflowHub.eu & \url{https://workflowhub.eu} \\
    & Dockstore & \url{https://dockstore.org} \\
    & nf-core & \url{https://nf-co.re} \\
    & Galaxy Toolshed & \url{https://toolshed.g2.bx.psu.edu} \\
    & bio.tools & \url{https://bio.tools} \\
    & PegasusHub.io & \url{https://pegasushub.io} \\
    
    \rowcolor[HTML]{F2F2F2}
    Hardware & OpenCL & \url{https://www.khronos.org/opencl} \\
    \rowcolor[HTML]{F2F2F2}
    & Rosetta 2 & \url{https://support.apple.com/en-gb/HT211861} \\
    
    Orchestration, & TOSCA & \url{http://docs.oasis-open.org/tosca/TOSCA/v1.0/TOSCA-v1.0.html} \\
    deployment & Maestro & \url{https://github.com/LLNL/maestrowf} \\
    and execution & & \\

    \rowcolor[HTML]{F2F2F2}
    Data stores & \multicolumn{2}{l}{RDBMS, NoSQL, NewSQL, Object Stores} \\

    \toprule
    \multicolumn{3}{l}{\bf Semantic Workflow Interoperability} \\
    \midrule
    
    Bioschemas & \multicolumn{2}{l}{\url{https://bioschemas.org/profiles/ComputationalWorkflow/1.0-RELEASE}} \\ 
    EDAM & \multicolumn{2}{l}{\url{http://edamontology.org}} \\
    Biocompute Object & \multicolumn{2}{l}{\url{https://www.biocomputeobject.org}} \\
    IWIR Metaworkflows & \multicolumn{2}{l}{\url{https://doi.org/10.1109/IWSG.2015.18}} \\
    wfdesc & \multicolumn{2}{l}{\url{https://wf4ever.github.io/ro/2016-01-28/wfdesc}} \\
    Petri nets & \multicolumn{2}{l}{\url{https://en.wikipedia.org/wiki/Petri_net}} \\
    
    \rowcolor[HTML]{F2F2F2}
    \multicolumn{3}{l}{Business process modelling: BPML, BPEL, BPMN} \\
    
    \multicolumn{3}{l}{Lightweight protocols formalization: protocols.io (\url{https://www.protocols.io/})} \\
    \multicolumn{3}{l}{TeSS training workflows (\url{https://tess.elixir-europe.org/workflows})} \\
    
    \rowcolor[HTML]{F2F2F2}
    W3C PROV & wfprov & \url{https://w3id.org/ro/2016-01-28/wfprov} \\
    \rowcolor[HTML]{F2F2F2}
    efforts & OPMW & \url{http://www.opmw.org/model/OPMW} \\
    \rowcolor[HTML]{F2F2F2}
    & P-plan & \url{http://www.opmw.org/model/p-plan} \\
    \rowcolor[HTML]{F2F2F2}
    & s-provenance & \url{https://github.com/KNMI/s-provenance} \\
    \rowcolor[HTML]{F2F2F2}
    & provone-v1-dev & \url{https://purl.dataone.org/provone-v1-dev} \\
    \rowcolor[HTML]{F2F2F2}
    & ISO 23494 & \url{https://doi.org/10.5281/zenodo.3901011} \\
    \rowcolor[HTML]{F2F2F2}
    & CWLProv & \url{https://doi.org/10.1093/gigascience/giz095} \\
    
    \toprule
    \multicolumn{3}{l}{\bf Organizational Workflow Interoperability} \\
    \midrule
    
    COMBINE & \multicolumn{2}{l}{\url{http://co.mbine.org/standards}} \\
    HL7 & \multicolumn{2}{l}{\url{https://www.hl7.org/fhir}} \\
    IVOA & \multicolumn{2}{l}{\url{https://ivoa.net}} \\
    GA4GH & \multicolumn{2}{l}{\url{https://www.ga4gh.org/genomic-data-toolkit}} \\
    OGF & \multicolumn{2}{l}{\url{https://www.ogf.org/ogf/doku.php/standards/standards}} \\
    
    \toprule
    \multicolumn{3}{l}{\bf Legal Workflow Interoperability} \\
    \midrule
    
    Data access & NIH Data Commons & \url{https://commonfund.nih.gov/commons} \\
    & caBIG & \url{https://en.wikipedia.org/wiki/CaBIG} \\
    
    \rowcolor[HTML]{F2F2F2}
    Funding & H2020 & \url{https://ec.europa.eu/programmes/horizon2020/en/home} \\
    \rowcolor[HTML]{F2F2F2}
    & EOSC & \url{https://eosc-portal.eu} \\

    \multicolumn{3}{l}{License compatibility Debian-Med (\url{https://www.debian.org/devel/debian-med})} \\
    
    \bottomrule
\end{longtable}
\endgroup

\newpage
\cleardoublepage\phantomsection\addcontentsline{toc}{section}{Appendix B: Participants and Contributors}
\section*{Appendix B: Participants and Contributors}
\label{appx:contributors}

\begin{longtable}[!h]{llp{10.5cm}}
\toprule
\textbf{First Name} & \textbf{Last Name} & \textbf{Affiliation} \\
\midrule
Dong    & Ahn         & Lawrence Livermore National Laboratory \\
\rowcolor[HTML]{F2F2F2} 
Ilkay   & Altintas    & UC San Diego \\
Yadu    & Babuji      & University of Chicago           \\
\rowcolor[HTML]{F2F2F2} 
Rosa M  & Badia       & Barcelona Supercomputing Center \\
Bartosz & Balis       & AGH University of Science and Technology              \\
\rowcolor[HTML]{F2F2F2} 
Vivien  & Bonazzi     & Deloitte     \\
Ryan   & Bunney      & International Centre for Radio Astronomy Research     \\
\rowcolor[HTML]{F2F2F2} 
Silvina & Caíno-Lores & University of Tennessee         \\
Scott   & Callaghan   & Southern California Earthquake Center  \\
\rowcolor[HTML]{F2F2F2} 
Henri   & Casanova    & University of Hawaii at Manoa   \\
Kyle    & Chard       & University of Chicago           \\
\rowcolor[HTML]{F2F2F2} 
Tainã & Coleman & University of Southern California   \\
Frederik   & Coppens & VIB / ELIXIR Belgium            \\
\rowcolor[HTML]{F2F2F2} 
Michael R. & Crusoe  & Common Workflow Language / VU Amsterdam / ELIXIR-NL \\
Kaushik    & De      & Univ. of Texas at Arlington     \\
\rowcolor[HTML]{F2F2F2} 
Daniel& de Oliveira    & Fluminense Federal University   \\
Frank & Di Natale & Lawrence Livermore National Lab \\
\rowcolor[HTML]{F2F2F2} 
Tu Mai Anh & Do      & University of Southern California   \\
Alexander  & Dunn    & Lawrence Berkeley Lab           \\
\rowcolor[HTML]{F2F2F2} 
Bjoern& Enders  & NERSC        \\
Thomas& Fahringer & University of Innsbruck         \\
\rowcolor[HTML]{F2F2F2} 
Rafael& Ferreira da Silva                  & University of Southern California   \\
Rosa& Filgueira & Heriot-Watt University          \\
\rowcolor[HTML]{F2F2F2} 
Anne& Fouilloux & University of Oslo, Norway      \\
Grigori    & Fursin  & OctoML.ai / cTuning foundation              \\
\rowcolor[HTML]{F2F2F2} 
Alban & Gaignard& CNRS, France \\
Alex& Ganose  & Lawrence Berkeley National Laboratory \\
\rowcolor[HTML]{F2F2F2} 
Daniel& Garijo  & Universidad Politécnica de Madrid   \\
Sandra& Gesing  & University of Notre Dame        \\
\rowcolor[HTML]{F2F2F2} 
Carole& Goble   & The University of Manchester    \\
Adil& Hasan   & Uninett Sigma2   \\
\rowcolor[HTML]{F2F2F2} 
Thuc& Hoang   & DOE/NNSA     \\
Dorran& Howell  & Tweag I/O    \\
\rowcolor[HTML]{F2F2F2} 
Milton& Hoz de Vila    & University of Leeds             \\
Sebastiaan & Huber   & EPFL         \\
\rowcolor[HTML]{F2F2F2} 
Shantenu   & Jha     & BNL / Rutgers\\
Rajesh& Kalyanam& Purdue University\\
\rowcolor[HTML]{F2F2F2} 
Daniel S.  & Katz    & University of Illinois          \\
Juergen    & Klenk   & Deloitte     \\
\rowcolor[HTML]{F2F2F2} 
Daniel& Laney   & Lawrence Livermore National Lab \\
Simone& Leo     & CRS4         \\
\rowcolor[HTML]{F2F2F2} 
Ulf & Leser   & Humboldt-Universität zu Berlin  \\
Douglas    & Lowe    & University of Manchester        \\
\rowcolor[HTML]{F2F2F2} 
Bertram    & Ludaescher& University of Illinois, Urbana-Champaign \\
Ketan & Maheshwari& Oak Ridge National Laboratory   \\
\rowcolor[HTML]{F2F2F2} 
Maciej& Malawski& AGH University of Science and Technology \\
Marta & Mattoso & Federal University of Rio de janeiro\\
\rowcolor[HTML]{F2F2F2} 
Rajiv & Mayani  & University of Southern California   \\
Mimi& McClure & NSF          \\
\rowcolor[HTML]{F2F2F2} 
Kshitij    & Mehta   & Oak Ridge National Laboratory   \\
Andre & Merzky  & Rutgers University              \\
\rowcolor[HTML]{F2F2F2} 
Edmund& Miller  & University of Texas at Dallas   \\
Todd& Munson  & Argonne      \\
\rowcolor[HTML]{F2F2F2} 
Jonathan   & Ozik    & Argonne National Laboratory     \\
Tapasya    & Patki   & Lawrence Livermore National Laboratory  \\
\rowcolor[HTML]{F2F2F2} 
Luc & Peterson& Lawrence Livermore National Laboratory  \\
Kitchka    & Petrova & AAAS-FPI     \\
\rowcolor[HTML]{F2F2F2} 
Loic& Pottier & University of Southern California   \\
Nicholas   & Pritchard & International Centre for Radio Astronomy Research (ICRAR) \\
\rowcolor[HTML]{F2F2F2} 
Lavanya    & Ramakrishnan   & LBNL         \\
Sashko& Ristov  & University of Innsbruck         \\
\rowcolor[HTML]{F2F2F2} 
Mehdi & Roozmeh & Karlsruhe Institute of Technology   \\
Stian & Soiland-Reyes  & The University of Manchester    \\
\rowcolor[HTML]{F2F2F2} 
Renan & Souza   & IBM Research \\
Alan& Sussman & National Science Foundation     \\
\rowcolor[HTML]{F2F2F2} 
Frédéric & Suter & CNRS/CC-IN2P3 \\
Douglas    & Thain   & University of Notre Dame        \\
\rowcolor[HTML]{F2F2F2} 
Benjamin   & Tovar   & University of Notre Dame        \\
Matteo & Turilli & Rutgers University \\
\rowcolor[HTML]{F2F2F2} 
Karan & Vahi    & University of Southern California   \\
Alvaro & Vidal-Torreira & Parallel Works \\
\rowcolor[HTML]{F2F2F2} 
Wendy & Whitcup & University of Southern California   \\
Michael    & Wilde   & Parallel Works Inc.             \\
\rowcolor[HTML]{F2F2F2} 
Alan& Williams& The University of Manchester, UK\\
Matthew    & Wolf    & Oak Ridge National Laboratory   \\
\rowcolor[HTML]{F2F2F2} 
Justin& Wozniak & Argonne National Laboratory     \\     

\bottomrule                                       
\end{longtable}

\newpage
\cleardoublepage\phantomsection\addcontentsline{toc}{section}{Appendix C: Agenda}
\section*{Appendix C: Agenda}
\label{appx:agenda}

\begin{table}[!h]
    \begin{tabular}{lp{12cm}}
        \toprule
        \textbf{Time} & \textbf{Topic} \\
        \midrule
        10:00-10:05am PDT & \makecell[l]{\textbf{Welcome and Introductions}\\\emph{\small Rafael Ferreira da Silva (University of Southern California)}} \\
        
        \rowcolor[HTML]{EEEEEE}
        10:05-10:15am PDT & \makecell[l]{\textbf{Workflows Community Summit Report and Key Findings}\\\emph{\small Henri Casanova (University of Hawai'i at Mano\=a)}} \\

        10:15-10:30am PDT & \textbf{Lightning Talks}
            \begin{compactitem}
                \item Topic 1: Definition of common workflow patterns and benchmarks
                \item[] \emph{\small Dan Laney (Lawrence Livermore National Laboratory)}
                \item[] \emph{\small Dorran Howell (Tweag I/O)}
                
                \item Topic 2: Identifying paths toward interoperability of workflow systems  
                \item[] \emph{\small Stian Soiland-Reyes (The University of Manchester)}
                \item[] \emph{\small Ilkay Altintas (UC San Diego)}

                \item Topic 3: Improving workflow systems' interface with legacy and emerging HPC software and hardware stacks
                \item[] \emph{\small Douglas Thain (University of Notre Dame)}
                \item[] \emph{\small Rosa Filgueira (Heriot-Watt University)}
                \vspace{-10pt}
            \end{compactitem} \\
            
        \rowcolor[HTML]{EEEEEE}
        10:30-10:35am PDT & \emph{5min Break (preparation for breakout sessions)} \\
        
        10:35-11:15am PDT & \textbf{Project Development Sessions}
        \begin{compactitem}
                \item Topic 1: Definition of common workflow patterns and benchmarks
                \item Topic 2: Identifying paths toward interoperability of workflow systems
                \item Topic 3: Improving workflow systems' interface with legacy and emerging HPC software and hardware stacks
                \vspace{-10pt}
            \end{compactitem} \\
            
        \rowcolor[HTML]{EEEEEE}
        11:15-11:30am PDT & \emph{15min Break} \\
        
        11:30-12:10pm PDT & \textbf{Project Roadmap Write Up Sessions}
        \begin{compactitem}
                \item Topic 1: Definition of common workflow patterns and benchmarks
                \item Topic 2: Identifying paths toward interoperability of workflow systems
                \item Topic 3: Improving workflow systems' interface with legacy and emerging HPC software and hardware stacks
                \vspace{-10pt}
            \end{compactitem} \\
        
        \rowcolor[HTML]{EEEEEE}
        12:10-12:25pm PDT & \emph{15min Break} \\
        
        12:25-12:40pm PDT & \textbf{Reports: Project Roadmap Sessions} \\

        \rowcolor[HTML]{EEEEEE}
        12:40-12:50pm PDT & \makecell[l]{\textbf{Final remarks and moving forward}\\\emph{\small Kyle Chard (University of Chicago)}} \\
        \bottomrule
    \end{tabular}
\end{table}

\end{document}